\documentclass[aps,showpacs,preprintnumbers,amsmath,amssymb,prb,twocolumn,superscriptaddress
]{revtex4-1}
\usepackage{txfonts}
\usepackage{graphicx}
\usepackage{dcolumn}
\usepackage{bm}
\usepackage{color}
\usepackage{subfigure}  
\def\comment#1{}

\def\slashchar#1{\setbox0=\hbox{$#1$}           
   \dimen0=\wd0                                 
   \setbox1=\hbox{/} \dimen1=\wd1               
   \ifdim\dimen0>\dimen1                        
      \rlap{\hbox to \dimen0{\hfil/\hfil}}      
      #1                                        
   \else                                        
      \rlap{\hbox to \dimen1{\hfil$#1$\hfil}}   
      /                                         
   \fi}                                         %

\def\sigmab{{\mbox{\boldmath $\sigma$}}}
\def\nablab{{\mbox{\boldmath $\nabla$}}}

\begin{document}

\title{Higgs Mechanism, Phase Transitions, and Anomalous Hall Effect in Three-Dimensional Topological Superconductors}

\author{Flavio S. Nogueira}
\affiliation{Institut f{\"u}r Theoretische Physik III, Ruhr-Universit\"at Bochum,
Universit\"atsstra\ss e 150, DE-44801 Bochum, Germany}
\affiliation{Institute for Theoretical Solid State Physics, IFW Dresden, Helmholtzstr. 20, 01069 Dresden, Germany}

\author{Asle Sudb{\o}}
\affiliation{Department of Physics, Norwegian University of
Science and Technology, N-7491 Trondheim, Norway}

\author{Ilya Eremin}
\affiliation{Institut f{\"u}r Theoretische Physik III, Ruhr-Universit\"at Bochum,
Universit\"atsstra\ss e 150, DE-44801 Bochum, Germany}
\affiliation{National University of Science and Technology ‘MISiS’, 119049 Moscow, Russian Federation}

\date{Received \today}

\begin{abstract}
We demonstrate that the Higgs mechanism in three-dimensional topological superconductors exhibits unique features
with experimentally observable consequences. The Higgs model we discuss has two superconducting components and
an axion-like magnetoelectric  term with the phase difference of the superconducting order parameters playing the
role of the axion field. Due to this additional term, quantum electromagnetic and phase  fluctuations lead to a
robust topologically non-trivial state that holds also in the presence of interactions. In this sense, we show that
the renormalization flow of the topologically nontrivial phase
cannot be continuously deformed into a topologically non-trivial one. One consequence of our analysis of quantum
critical fluctuations, is the possibility of having a first-order phase transition in the bulk and a second-order phase
transition on the surface.  We also explore another consequence of the axionic Higgs electrodynamics, namely, the
anomalous Hall effect. In the low frequency London regime an anomalous Hall effect is induced in the presence of an applied electric field
parallel to the surface. This anomalous Hall current is induced by a Lorentz-like force arising from the axion term,
and it  involves the relative superfluid velocity of the superconducting components. The anomalous Hall current has
a negative sign, a situation reminiscent of, but quite distinct in physical origin from the anomalous Hall effect observed
in high-$T_c$ superconductors. In contrast to the latter, the anomalous Hall effect in topological superconductors
is non-dissipative and occurs in the absence of vortices.
\end{abstract}

\maketitle

\section{Introduction}

Topological solid states of matter \cite{Hasan-Kane-RMP,Zhang-RMP-2011} have the property that their bulk states are gapped,
while states at the boundaries are gapless and protected by some discrete quantum symmetry. The topological aspect
emerges when considering the transport properties of the boundary states, where the transport current happens to also be a
topological current. The most well established topological solid states of matter are topological insulators (TIs), which are
gapped in the bulk and have helical (or chiral) gapless states at the boundaries which are
protected by time-reversal symmetry. Helical here means that the electronic spin is locked
to momentum due to strong spin-orbit coupling. Thus, the boundary states have an helicity
determined by the eigenvalues of $\sigmab\cdot{\bf k}/k$ at each boundary. Topological
insulators have been predicted to exist \cite{Prediction-TI} and confirmed experimentally in subsequent papers \cite{Experiments-TI}. Although many of the materials investigated
experimentally are not perfect insulators in the bulk, the observed boundary helical
states are robust features of these materials.

There exists another predicted topological solid state of matter, namely topological superconductors (TSCs) \cite{Hasan-Kane-RMP,Zhang-RMP-2011}, where the experimental situation is less clear.
TSCs  follow a symmetry classification scheme closely related to TIs, as far as Hamiltonians of the
Bogoliubov-de Gennes type are concerned.\cite{Classif} Just like TIs, TSCs have gapped states
in the bulk and symmetry-protected gapless states at the boundaries. Unlike TIs, in TSCs the
$U(1)$ symmetry is broken, either spontaneously or by proximity effect. The gapless boundary
states are Majorana fermions, which are fermionic particles that are their own anti-particles. In order to support such states at the boundaries, the topological superconductivity must feature a
$p$-wave type of gap. Particle-hole is the underlying symmetry protecting the boundary Majorana states. In one dimension, a paradigmatic
simple model for topological superconductivity has been proposed by Kitaev \cite{Kitaev} where the Majorana zero-energy states
live at the ends of a quantum wire. An experimental way of realizing a superconducting state in a quantum wire is by proximity
effect. In this case a semiconducting wire with strong spin-orbit coupling is deposited on the surface of an $s$-wave superconductor in the presence of an external perpendicular magnetic
field. Then by proximity effect $p$-wave like superconductivity is induced on the wire for a certain range of parameters.\cite{Alicea-2011,Tewari-review} There are some experimental signatures of Majorana modes in Indium antimonide nanowires in contact with normal and superconducting electrodes.\cite{Mourik} More recently, strong support for Majorana boundary zero modes have been reported in an experiment with a ferromagnetic chain of Iron fabricated
on the surface of lead.\cite{Majo-Fe-chain}

Three-dimensional (3D) TSCs have also been discussed theoretically, in particular focusing on vortex physics
\cite{Hasan-Kane-RMP,Zhang-RMP-2011} and possible topological phase transitions.\cite{Hosur-2011} A distinctive feature of both
three-dimensional TIs and TSCs with respect to their non-topological counterparts is the topological magnetoelectric response induced
by a mechanism similar to the
so-called chiral anomaly.\cite{chiral-anomaly} When fermions in topological materials interact with the electromagnetic field, a Berry
phase mixing electric and magnetic fields is induced.\cite{Qi-2008,Qi-Witten-Zhang} In TIs this occurs due to strong spin-orbit coupling that locks spin to momentum. The resulting Berry
phases combine in the form of a so-called axion term, which is a magnetoelectric term
$\sim {\bf E}\cdot{\bf B}$ with a periodic field, $\theta$, appearing as a
coefficient.\cite{Qi-2008}  This coefficient corresponds to a topological invariant implied
by the Chern number of the band structure. If TR symmetry is preserved, there are only two
possible values for $\theta$ in a 3D insulator, namely, $\theta=0$ and $\pi$, the
former value corresponding to a topologically trivial insulator.\cite{Qi-2008}
In the case of TSCs a topological magnetoelectric contribution also arises, but
now $\theta$ corresponds to the phase difference between order parameters of opposite
chirality.\cite{Qi-Witten-Zhang} Recently, axionic superconductivity has been also
discussed in the context of doped narrow-gap semiconductors.\cite{Roy-2014}

The 3D TSC constructed by Qi {\it et al.} \cite{Qi-Witten-Zhang} should actually correspond to an interacting topological state of matter, going beyond the the symmetry classification of
topological non-interacting theories for insulators and superconductors.\cite{Classif}
According to the standard symmetry classification and an argument following from the
gravitational anomaly,\cite{Ryu} 3D TSCs having TR invariance are classified by an integer topological invariant, belonging
to the DIII class in the free fermion classification table.\cite{Classif}
Thus, unlike TR invariant TIs, the electromagnetic axionic response
of 3D TSCs would involve an axion field having values $\theta=0$ or $\pi$, mod $2\pi$.
However, as pointed out in Ref. \onlinecite{Qi-Witten-Zhang}, the axion in TSCs is a
dynamical phase variable associated with the superconducting order parameter, or Higgs
field in the language of quantum field theory. Furthermore, the Meissner effect gives a
mass to the photon via the Higgs mechanism. This is an important difference
with respect to the axion electrodynamics of TIs, where no U(1) symmetry is broken and the
gauge field remains gapless. Furthermore, nontrivial Chern numbers are
associated with the different phases. For the simpler case featuring two Weyl fermions with
opposite helicity, we have opposite Chern numbers, leading to  a topological invariant given
by the sign of the gap amplitude.\cite{Qi-Witten-Zhang}  Thus, the arguments in Ref.  \onlinecite{Qi-Witten-Zhang} seems to
point out to a $Z_2$ classification. However, since the relevant symmetry in the problem is $U(1)\times Z_2$,
it has been shown recently \cite{Vishwanath-PRX-Z16,Wang-Senthil-Z16}
that in topological superconductors with TR symmetry the $\mathbb{Z}$ classification is reduced to
$Z_{16}$.

In this paper, we investigate the Higgs mechanism and anomalous Hall effect of three-dimensional TSCs within the model introduced recently in Ref.  \onlinecite{Qi-Witten-Zhang}. In the
simplest case, the model features two superconducting order parameters associated with left
and right fermion chiralities interacting with the electromagnetic field, and a topological
magnetoelectric term. We will show that quantum fluctuations in such a SC imply an
interacting topologically non-trivial phase that cannot be continuously deformed
into the interacting topologically trivial one. Our claim is substantiated by a renormalization
group analysis. This result does not follow from the classical Lagrangian of the system derived
earlier in Ref.\onlinecite{Qi-Witten-Zhang}, and is a purely quantum effect involving the interaction between photons and Higgs fields. We show that due to this behavior,
the type I regime of the TSC features a first-order phase transition in the bulk and a
second-order phase transition on the surface. This distinguishes a TSC from a topologically
trivial superconductor in the type I regime. The latter would exhibit a first-order phase
transition in the bulk as well as on the surface, provided quantum fluctuations are accounted
for \cite{Halperin-Lubensky-Ma} (this point will be discussed in detail in Section III).

Another consequence of the topological magnetoelectric term is the occurrence of an anomalous
Hall effect when an electric field is applied parallel to the surface of a TSC. Due to the magnetoelectric effect, a transverse current is generated from a Lorentz-like force involving
the relative superfluid velocity $\sim\nablab\theta$ and the applied electric field. The generated transverse current is negative, a situation reminiscent of the anomalous Hall effect in superconductors \cite{Josephson-Hall-effect}, and observed  high-$T_c$ cuprate
superconductors.\cite{Anom-Hall-Effect-high-Tc} However, in the latter case the anomalous
Hall effect occurs due to vortex motion induced by the Faraday law, and is typically a very
small effect. Furthermore, in this case the Lorentz force acts directly on the vortex core,
and therefore on the normal components of the superconductor. For this reason, it
automatically leads to dissipation. In the case of three-dimensional TSCs, on the other hand,
the anomalous Hall effect occurs even in absence of vortices and is induced solely by an
external electric field via the magnetoelectric effect. Thus, a {\it dissipationless
anomalous Hall current is generated on the surface}.

The plan of the paper is as follows. In Section II we discuss how the Higgs mechanism works
in a topological superconductor. We will show that after the phases are integrated out,
interactions between the photons automatically arise due to the axion term. This is in
contrast to the ordinary Higg mechanism, where the phases can be trivially integrated
out by a gauge transformation. In Section III we discuss the role of quantum fluctuations
and derive the effective potential on the surface and  the renormalization group (RG)
equations. We show that the RG equations of the topological superconductor cannot be
connected to the ones of a topologically trivial superconductor. This is shown to occur
as a direct consequence of the axion term. Finally, in Section IV we discuss the
dissipationless variant of the anomalous Hall effect using the London limit of
topological superconductors.

\section{Higgs mechanism in three-dimensional topological superconductors}

The effective Lagrangian for a three-dimensional TSC featuring two Fermi surfaces is
given by \cite{Qi-Witten-Zhang},
\begin{eqnarray}
\label{Eq:L-Higgs}
 {\cal L}_{\rm eff}&=&\frac{e^2\theta}{32\pi^2}\epsilon^{\mu\nu\sigma\tau} F_{\mu\nu}F_{\sigma\tau}
-\frac1{4}F_{\mu\nu}F^{\mu\nu}\nonumber\\
&+&\sum_{i=L,R}\left[|(\partial_\mu-qA_\mu)\phi_i|^2-m^2|\phi_i|^2\right]\nonumber\\
&-&\frac{u}{2}(|\phi_L|^2+|\phi_R|^2)^2+2J(\phi_L^*\phi_R+\phi_R^*\phi_L),
\end{eqnarray}
where $q=2e$ is the charge of the condensate. The Greek indices run from 0 to 3 and $F_{\mu\nu}=\partial_\mu A_\nu-\partial_\nu A_\mu$, with $(A_\mu)=(A_0,{\bf A})$.
The Lagrangian (\ref{Eq:L-Higgs}) corresponds to an abelian Higgs model with a two-component scalar field $(\phi_L,\phi_R)$, where $R,L$ denote "right" and "left" chiralities of the two components of the scalar condensate matter field, minimally coupled to the gauge-field $(A_\mu)$. In contrast to the standard Higgs model, Eq. (\ref{Eq:L-Higgs}) features a so-called axion term \cite{Witten,Wilczek}, which is the first term appearing in the Lagrangian above. The term is topological in nature and contains a scalar field (the axion)
$\theta=\theta_L-\theta_R$, where $\theta_L$ and $\theta_R$ are the phases of $\phi_L$ and
$\phi_R$, respectively, i.e. $\phi_i = |\phi_i| \exp(i \theta_i)$. In terms of electric and magnetic field components, the toplogical term reads
$e^2\theta{\bf E}\cdot{\bf B}/(4\pi^2)$, which is precisely the magnetoelectric form mentioned
in the introductory paragraphs. A Josephson coupling term $\propto \phi_L^*\phi_R+\phi_R^*\phi_L$  accounts for the interference between the two superconducting order parameter fields. This is
a characteristic feature in superconductors with two or more components of the order parameter,
and is absent only if prohibited by symmetries of the problem .
\cite{Babaev-Sudbo-2005,Sudbo-2005} Below, we show that the Josephson coupling is generated
by fluctuations, and therefore it is legitimate to include such a term
from the very beginning in the Lagrangian. Furthermore, the Josephson coupling is important
for tuning between topologically trivial and nontrivial phases. In fact,  a simple mean-field analysis shows that for $J<0$ the Josephson coupling implies $\theta=0$, yielding a topologically trivial superconductor. For $J>0$, on the other hand, $\theta$ is locked to $\pi$, thus leading
to a topologically non-trivial superconducting ground state. Since $\theta$ is periodic,
$\theta=\pi$ corresponds to a situation where the time-reversal (TR) symmetry is preserved
\cite{Qi-Witten-Zhang}. Thus, at the mean-field level, $J=0$ separates a topologically
trivial ground state from a non-trivial one. Thus, varying $J$ from positive to
negative values induces a topological quantum phase transition.

In the $U(1)$ Higgs mechanism the phases disappear from the spectrum due to spontaneous breaking of the local $U(1)$ symmetry, being transmuted into the longitudinal mode for the photon, which becomes gapped. Thus, only amplitude modes remain in the spectrum of scalar particles. The Higgs mechanism is equivalent to integrating out the phase degrees of freedom, which in the case of the Higgs model automatically leads to a massive gauge particle. This point of view of integrating out the phases
is particularly appealing in the case where a Josephson coupling is present. However, additional
non-linearities arise in the presence of the axion term. To see this, let us first consider the Higgs mechanism in Eq. (\ref{Eq:L-Higgs}) for the case where the axion term is absent. In this
case we can simply write $\phi_j=\rho_j e^{i\theta_j}/\sqrt{2}$, $j=L,R$ and
make the shift,
\begin{equation}
\label{Eq:shift}
 A_\mu\to A_\mu+\frac{1}{q}\left(\frac{\rho_L^2\partial_\mu\theta_L+\rho_R^2\partial_\mu\theta_R}{\rho_L^2+\rho_R^2}\right),
\end{equation}
which yields,
\begin{eqnarray}
{\cal L}_{\rm eff}&=&\frac{e^2\theta}{32\pi^2}\epsilon^{\mu\nu\sigma\tau} F_{\mu\nu}F_{\sigma\tau}
-\frac1{4}F_{\mu\nu}F^{\mu\nu}+\frac{q^2(\rho_L^2+\rho_R^2)}{2}A_\mu^2\nonumber\\
&+&\frac{\rho_L^2\rho_R^2}{2(\rho_L^2+\rho_R^2)}(\partial_\mu\theta)^2+J\rho_L\rho_R\cos\theta
\nonumber\\
&-&\frac{m^2}{2}(\rho_L^2+\rho_R^2)-\frac{u}{8}(\rho_L^2+\rho_R^2)^2.
\end{eqnarray}
Note that in absence of the axion term and for the particular situation of a single scalar field component, i.e., $\rho_L=\rho$ and $\rho_R=0$, the above effective Lagrangian trivially reduces
to the usual Lagrangian for the Higgs model in the unitary gauge. In this particular case the Lagrangian is independent of the phase.  For the case relevant to us here,  where two scalar
fields are present, there is a term $\sim (\partial_\mu\theta)^2$ remaining. Thus, in the absence of Josephson coupling there is still a massless (Goldstone) mode present in the spectrum. This occurs because there are two Higgs fields and one Abelian gauge field. Thus, it is only possible to gauge away one phase degree of freedom. The phase $\theta$ would not couple directly to the gauge field
in the absence of the axion term. Due to the axion term, integrating out the phases generate direct interactions between photons, even if the amplitudes are assumed to be uniform. For the topological phase of the system occurring for $J>0$, we have to integrate out the lowest order Gaussian phase fluctuations around $\theta=\pi$. In general, this renders the induced photon-photon interaction being non-local. If  the amplitudes are uniform, we obtain the effective Lagrangian,
\begin{eqnarray}
\label{Eq:Higgs-J>0}
&&{\cal L}_{\rm Higgs}|_{J>0}=-\frac{1}{4}F^2+\frac{q^2(\rho_L^2+\rho_R^2)}{2}A^2+
\frac{e^2}{32\pi}\epsilon^{\mu\nu\sigma\tau} F_{\mu\nu}F_{\sigma\tau}
\nonumber\\
&-&\frac{1}{2}\left(\frac{e^2}{16\pi^2}\right)^2\left(\frac{\rho_L^2+\rho_R^2}{\rho_L^2\rho_R^2}
\right)\nonumber\\
&\times&\int d^4x'V(x-x')\epsilon_{\mu\nu\lambda\rho}\epsilon^{\alpha\beta\gamma\delta}
F^{\mu\nu}(x)F^{\lambda\rho}(x)F_{\alpha\beta}(x')F_{\gamma\delta}(x')
\nonumber\\
&-&J\rho_L\rho_R-\frac{m^2}{2}(\rho_L^2+\rho_R^2)-\frac{u}{8}(\rho_L^2+\rho_R^2)^2,
\end{eqnarray}
where
\begin{equation}
 V(x)=\int\frac{d^4p}{(2\pi)^4}\frac{e^{ip\cdot x}}{p^2+m_\theta^2},
\end{equation}
with $m_\theta^2=J(\rho_L/\rho_R+\rho_R/\rho_L)$. In the long wavelength regime the induced photon interaction is strongly screened by the axion. Hence, we have  $V(x-x')\approx m_\theta^{-2}\delta^4(x-x')$.
The resulting photon-photon interaction simplifies and we obtain\cite{Note2}
\begin{eqnarray}
\label{Eq:Higgs-J>0-1}
&&{\cal L}_{\rm Higgs}|_{J>0}=-\frac{1}{4}F^2+\frac{q^2(\rho_L^2+\rho_R^2)}{2}A^2+
\frac{e^2}{32\pi}\epsilon^{\mu\nu\sigma\tau} F_{\mu\nu}F_{\sigma\tau}
\nonumber\\
&+&\frac{1}{2J\rho_L\rho_R}\left(\frac{e^2}{16\pi^2}\right)^2\det(F_{\mu\nu})\nonumber\\
&-&J\rho_L\rho_R-\frac{m^2}{2}(\rho_L^2+\rho_R^2)-\frac{u}{8}(\rho_L^2+\rho_R^2)^2,
\end{eqnarray}
where $\det(F_{\mu\nu})=({\bf E}\cdot{\bf B})^2$.

\section{Quantum fluctuations}

\subsection{Important vanishing of a Feynman diagram}

\begin{figure}
 \centering
 \includegraphics[width=8cm]{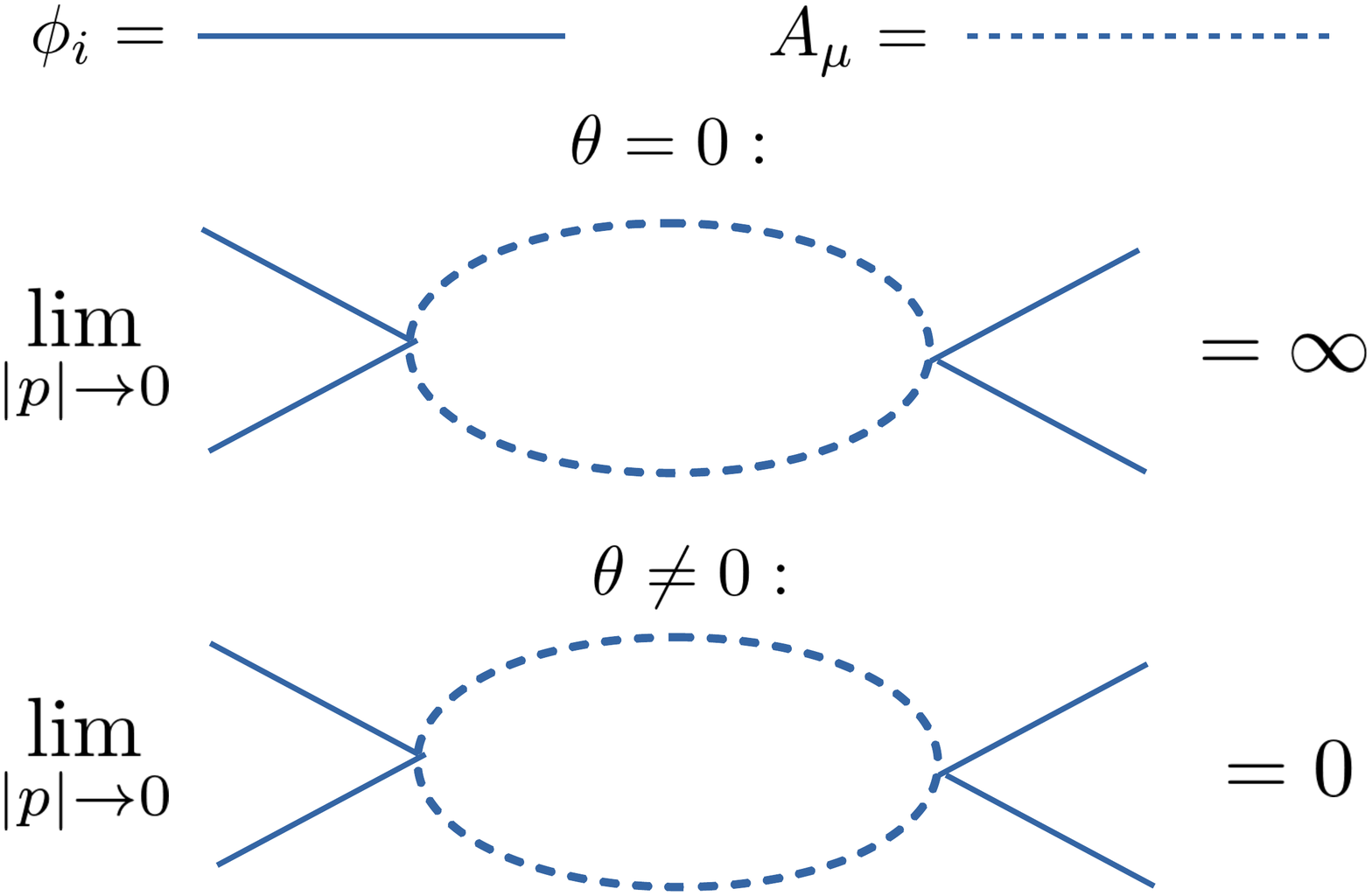}
 \caption{Difference in behavior for surface photon-mediated Higgs scattering at the critical
 point in
 the long wavelength limit. For the topologically trivial case ($\theta=0$) the corresponding
 Feynman
 diagram diverges. On the other hand, for the topologically non-trivial case ($\theta\neq 0$), the same diagram vanishes. This shows beyond the non-interacting regime that a topological superconductor cannot be continuously deformed into a topologically non-trivial
 one in the long wavelength limit.}
 \label{Fig:fish}
\end{figure}

We now turn to a crucial aspect of the topological phase with respect to the surface states. It turns out that in the presence of quantum fluctuations, the
topological surface states cannot be continuously deformed into topologically trivial ones
in the long wavelength limit when crossing the critical point. To see, this let us
assume that $\theta$ is uniform on the surfaces, with $\theta=\pi$ for the
TR invariant case. Note that $\theta_L$ and $\theta_R$ are still
allowed to fluctuate, with
$\theta_L=\theta/2+\delta\theta_L$ and $\theta_R=-\theta/2+\delta\theta_R$.
Since  $\epsilon^{\mu\nu\sigma\tau} F_{\mu\nu}F_{\sigma\tau}=2\partial^\mu (\epsilon_{\mu\nu\lambda\rho}A^\nu F^{\lambda\rho})$, each surface contains a Chern-Simons (CS)
term.\cite{CS} Thus, assuming two surfaces perpendicular to the $z$-axis, we find that
the imaginary time photon propagator at any surface is given by
\begin{equation}
 \label{Eq:photon-prop}
 \Delta_{\mu\nu}^\pm(p)=\frac{p^2+m_A^2}{\Delta(p^2)}\left[\delta_{\mu\nu}-\frac{p_\mu p_\nu}{p^2}
 \pm\frac{M_\theta}{p^2+m_A^2}\epsilon_{\mu\nu\lambda}p^\lambda\right],
\end{equation}
where $m_A^2=q^2(\rho_L^2+\rho_R^2)$, $M_\theta=e^2\theta/(8\pi^2)$,
$\Delta(p^2)=(p^2+m_A^2)^2+M_\theta^2p^2$, and
the $\pm$ sign is chosen depending on which surface one is referring to. In Eq. (\ref{Eq:photon-prop}) the transverse gauge
has been fixed. At the phase transition to the normal state where $m_A^2\to 0$, the Feynman diagram shown
in Fig. \ref{Fig:fish}, which is associated with Higgs scattering mediated by photons, behaves very differently at long
wavelengths ($|p|\to 0$), depending on whether $\theta=0$ (topologically trivial) or $\theta\neq 0$ (topologically non-trivial). Namely, for all $\theta\neq 0$ we have,
\begin{equation}
 \label{Eq:fish-diagram}
 \lim_{|p|\to 0}\lim_{m_A^2\to 0}\int\frac{d^3q}{(2\pi)^3}\Delta_{\mu\nu}^\pm(p+q)\Delta_{\nu\mu}^\pm(q)=0,
\end{equation}
whereas the result is divergent for $\theta=0$.
Thus, after other one-loop scattering amplitudes are taken into account to obtain the full four-Higgs vertex, we see that at the critical point the topological field theory cannot be continuously deformed into a topologically trivial one.
This statement holds trivially for topological Bogoliubov-de Gennes superconductors. Here,
we have shown that it also holds in the presence of quantum fluctuations in an interacting
system, beyond the Bogoliubov-de Gennes picture. This occurs because the photon is
topologically gapped, despite the vanishing of the Meissner gap ($m_A=0$).  Note that this
result is due to the topological character of the axion term, and not due to a symmetry
protection. Indeed, since  the diagram of Fig. \ref{Fig:fish}
vanishes for any $\theta\neq 0$, TR invariance is not required.  Thus, in this context the topologically non-trivial
phase simply corresponds to the case where the axion term is nonzero.

A continuous deformation to the topologically trivial phase can be done in the Higgs phase,
where there are no gapless modes. At the critical point such a continuous deformation is not possible. Thus, quantum critical fluctuations in this system will govern topologically
stable universal behavior in physical quantities. This has important implications for
critical exponents and amplitude ratios. Indeed, as we will see, the vanishing  of the
diagram shown in Fig. \ref{Fig:fish} changes significantly the renormalization group
(RG) $\beta$ function associated to the interaction vertex between scalar fields.

\subsection{One-loop effective action on the surface}

As is well known, the one-loop effective action is more easily obtained by integrating out the quadratic fluctuations of the
scalar fields and gauge fields.\cite{Dolan-Jackiw-1974} We will assume that the magnitudes
of both $\phi_L$ and $\phi_R$ have the same expectation value in the broken symmetry phase.
Thus, we write $\phi_L=e^{i\theta/2}\varphi+\tilde \phi_L$ and
$\phi_R=e^{-i\theta/2}\varphi +\tilde \phi_R$, where $\varphi$ is uniform and
$\tilde \phi_i$ represents the fluctuation around $\langle\phi_i\rangle$. The effective
action is therefore written in the form,
\begin{widetext}
\begin{equation}
\label{Eq:Gauss-flucts}
S_{\rm eff}^{\rm 1-loop}=S_{\rm eff}^0+\frac{1}{2}\int d^3x\int d^3x'\left[\Phi^\dagger(x)M(x-x';\varphi)\Phi(x')
+A_\mu(x) M_{\mu\nu}(x-x';\varphi)A_\nu(x')\right],
\end{equation}
where
\begin{equation}
S_{\rm eff}^0=2V[(m^2-2J\cos\theta)|\varphi|^2+u|\varphi|^4],
\end{equation}
with $V$ being the (infinite) volume of three-dimensional spacetime,
and
$\Phi^\dagger=[\tilde \phi_L^*~~\tilde \phi_R^*~~\tilde \phi_L~~\tilde \phi_R]$. The matrices
$M(x-x';\varphi)$ and $M_{\mu\nu}(x-x';\varphi)$ are given in momentum space by,
\begin{equation}
\label{Eq:M}
M(p,\varphi)=
\left[
\begin{array}{cc}
(p^2+m^2+3u|\varphi|^2)I+(u|\varphi|^2e^{i\sigma_z\theta}-2J)\sigma_x & u|\varphi|^2(e^{i\theta\sigma_z}+\sigma_x)\\
\noalign{\medskip}
u|\varphi|^2(e^{-i\theta\sigma_z}+\sigma_x) & (p^2+m^2+3u|\varphi|^2)I+(u|\varphi|^2e^{-i\sigma_z\theta}-2J)\sigma_x
\end{array}
\right],
\end{equation}
where $I$ is a $2 \times 2$ identity matrix, while $\sigma_x$ and $\sigma_z$  are Pauli matrices, and
\end{widetext}
\begin{equation}
M_{\mu\nu}(p,\varphi)=(p^2+4q^2|\varphi|^2)\delta_{\mu\nu}-p_\mu p_\nu-M_\theta\epsilon_{\mu\nu\lambda}p_\lambda.
\end{equation}
From Eq. (\ref{Eq:M}), we see that a correction to the Josephson coupling has been generated by fluctuations. It has the form,
$u|\varphi|^2(e ^{i\theta}\tilde \phi_L^*\tilde \phi_R+e^{-i\theta}\tilde \phi_R^*\tilde \phi_L)$. This term is generated even if $J=0$, which means that including a Josephson coupling in
the Lagrangian is a physically reasonable assumption. This should be expected, since
fluctuations will necessarily lead to an overlap between the two complex field components.
Note that this result is valid in general for any two-component superconductor and is not
restricted to the topological one being considered here.

Integrating out the fluctuations in Eq. (\ref{Eq:Gauss-flucts}), we obtain,
\begin{equation}
e^{-V U(|\varphi|,\theta)}=\int{\cal D}A_\mu{\cal D}\Phi^\dagger{\cal D}\Phi e^{-S_{\rm eff}^{\rm 1-loop}(A_\mu,\Phi^\dagger,\Phi)},
\end{equation}
where $V$ is the three-dimensional spacetime volume and $U(|\varphi|,\theta)$ is the effective potential given by,
\begin{equation}
 U(|\varphi|,\theta)=\frac{1}{2V}\left[{\rm Tr}\ln M_{\mu\nu}+{\rm Tr}\ln M\right].
\end{equation}
The tracelogs can be written in more explicit form,
\begin{eqnarray}
\label{Eq:U}
&& U(|\varphi|,\theta)=2[(m^2-2J\cos\theta)|\varphi|^2+u|\varphi|^4]
\nonumber\\
&+&\frac{1}{2}\sum_{\sigma=\pm}\int\frac{d^3p}{(2\pi)^3}\left\{\ln\left[p^2+M_\sigma^2(|\varphi|,\theta)\right]
\right.\nonumber\\
&+&\left.\ln\left[p^2+M_{1\sigma}^2(|\varphi|,\theta)\right]+\ln\left[p^2+M_{2\sigma}^2(|\varphi|,\theta)\right]
\right\},
\end{eqnarray}
where,
\begin{equation}
\label{Eq:M+-}
M_\pm^2(|\varphi|,\theta)=2q^2|\varphi|^2+\frac{M_\theta^2}{2}\pm\frac{|M_\theta|}{2}
\sqrt{8q^2|\varphi|^2+M_\theta^2},
\end{equation}
\begin{equation}
 M_{1\pm}^2(|\varphi|,\theta)=m^2+2u|\varphi|^2\pm2 J,
\end{equation}
\begin{eqnarray}
&& M_{2\pm}^2(|\varphi|,\theta)=m^2+4u|\varphi|^2
\nonumber\\
&\pm&2\sqrt{J^2+u^2|\varphi|^4-2Ju\cos\theta|\varphi|^2}.
\end{eqnarray}
From this mass spectrum underlying the effective potential, we recognize $p^2+M_{1\pm}^2$
and $p^2+M_{2-}^2$ as the would-be Goldstone modes from the regime $J=0$,
corresponding to the absence of the Josephson coupling. Note that the corresponding uncharged
system in absence of Josephson coupling features three Goldstone modes. This is to be expected, since the system in this case would be $O(4)$ invariant. As usual, in the charged system there
are only gapped modes, even in absence of Josephson coupling, as required by the Higgs mechanism.

The momentum integral in Eq. (\ref{Eq:U}) can be easily evaluated using an ultraviolet cutoff $\Lambda$,
\begin{eqnarray}
\label{Eq:U-1}
&& U(|\varphi|,\theta) = U_0
+ 2\left[\left(m^2-2J\cos\theta\right)|\varphi|^2+u|\varphi|^4\right]
\nonumber\\
&-&\frac{1}{12\pi}\sum_{\sigma=\pm}\left[|M_\sigma(|\varphi|,\theta)|^3+|M_{1\sigma}(|\varphi|,\theta)|^3
\right.\nonumber\\
&+& \left. |M_{2\sigma}(|\varphi|,\theta)|^3\right],
\end{eqnarray}
where $U_0$ is a field-independent (vacuum) contribution. We have absorbed a term
$\Lambda(3u+q^2)/(2\pi^2)$ in $m^2$, since this contribution just corresponds to tadpole
diagrams. Note that this simple renormalization does not actually affect the other terms,
since the error committed by doing this is of higher order than one-loop. From the
effective potential above we see that the diagram of Fig.  \ref{Fig:fish}
can only arise from the sum $\sum_{\sigma=\pm}|M_\sigma(|\varphi|,\theta)|^3$, with $M_\sigma^2(|\varphi|,\theta)$ given in Eq. (\ref{Eq:M+-}). An expansion in powers of $|\varphi|$ for
$\theta\neq 0$ clearly shows that the diagram shown in the figure indeed vanishes
when $\theta\neq 0$. However, the contributions $q^{2n}|\varphi|^{2n}$ for $n\geq 4$
become singular {\it for all} $\theta$. We emphasize that this is only true when the
topological magnetoelectric term is nonzero. On the other hand, if $\theta=0$, we obtain,
\begin{equation}
\label{Eq:cubic}
\sum_{\sigma=\pm}|M_\sigma(|\varphi|,\theta=0)|^3=8q^3|\varphi|^3,
\end{equation}
meaning that the effective potential is non-analytic as a function of $q ^2|\varphi|^2$.
Albeit simple, this is actually a non-perturbative result, since it cannot be obtained as a
power series in $q^2$. The absence of a power series involving terms $q^{2n}|\varphi|^{2n}$
reflects the divergence of the diagram shown in Fig.\ref{Fig:fish} when $\theta=0$ for
vanishing external momenta. In this case a meaningful perturbative evaluation of the vertex
function can only be done for nonzero external momenta, a fact related to the infrared
divergence arising from a massless photon. Note that in contrast to the $\theta\neq 0$
case, the expansion is singular in the infrared for all $n\geq 2$, rather than
for $n\geq 4$. However, the singularities in the case of a topologically massive photon
is not a problem, since they correspond to interactions that are irrelevant in an RG sense.
For instance, this type of infrared singularity in higher order vertices would also occur
in a simple $\varphi^4$ Landau theory. Having $\theta\neq 0$ turns the photon topologically
massive without spoiling gauge invariance.\cite{CS}  Thus, we can interpret the cubic
contribution arising in the limit of vanishing $\theta$ as a consequence of resumming up
all the one-loop infrared divergent diagrams containing only internal photon lines. This
leads to a non-analytic contribution to the effective potential. As is well-known,
summing up these contributions in $d=3+1$ in the case of standard scalar electrodynamics
leads to a logarithmic term $\sim q^4|\varphi|^4\ln |\varphi|^2$ in the effective
potential,\cite{Coleman-Weinberg} while the cubic term has been obtained in the context
of Ginzburg-Landau superconductors for $d=3+0$ (i.e., scalar electrodynamics in $d=2+1$
and imaginary time).\cite{Halperin-Lubensky-Ma} In both cases these one-loop photon
contributions lead to a fluctuation-induced first-order phase transition. However, it
was later shown that this result is valid only in the type I regime, while in the
type II regime a second-order phase transition in the so-called inverted 3D XY universality
class arises.\cite{Dasgupta}

In view of the above discussion, it is of interest to investigate the character of the superconducting phase transition on the surface of a topological superconductor. The
theory features two ingredients that are not present in the previous analysis of
fluctuation-induced first-order phase transitions by Halperin {\it et al}.
\cite{Halperin-Lubensky-Ma} These are the Josephson coupling between the scalar field
components, and the CS term. The case with Josephson coupling and in absence of CS
term has been examined in the London limit (i.e., in the strong type II regime)
in Ref. \onlinecite{Sudbo-2005}. For this case it has been shown by means of exact
duality arguments that a two-component superconductor with Josephson coupling exhibits
a phase transition in the 3D XY universality class.\cite{Sudbo-2005} On the other hand,
when the Josephson coupling is absent, it has been shown using renormalization group (RG)
methods in Ref. \onlinecite{deCalan} that for a large enough CS coupling (i.e., $M_\theta$
in our notation), the first-order phase transition is turned into a second-order one.
Regarding the result in the London limit without CS term, we note that our analysis in
this Section is being done in the type I regime, since amplitude fluctuations play an
important role in the calculations above. Indeed, the theory with $\theta= 0$ yields a
fluctuation induced first-order phase transition.  

\subsection{Renormalization group analysis}

It is a well-known fact that the CS term does not renormalize.\cite{Coleman-Hill} This is a consequence of the topological nature of the CS term. Indeed, since it is independent of the
metric, it does not change under scale transformations. Thus,  using this result and the
invariance of the effective action under renormalization, we obtain,
\begin{equation}
M_\theta\epsilon_{\mu\nu\lambda}A^\mu\partial^\nu A^\lambda=M_{\theta,r}\epsilon_{\mu\nu\lambda}A_r^\mu\partial^\nu A_r^\lambda,
\end{equation}
where $A_r^\mu=Z_A^{-1/2}A^\mu$ is the renormalized gauge field with the corresponding
wavefunction renormalization, $Z_A$, and $M_{\theta,r}$ is the renormalized topological
mass. From the above equation it follows that $M_{\theta,r}=Z_AM_\theta$.
Since gauge invariance implies the renormalization $q^2_r=Z_Aq ^2$,
\cite{Coleman-Weinberg,Halperin-Lubensky-Ma}  which is easily obtained
from the vacuum polarization, it follows that $\theta$ is a renormalization
group invariant,
\begin{equation}
\label{Eq:RG-inv}
\frac{d\theta}{dl}=0,
\end{equation}
where $l=\ln(m_r/\Lambda)$ is a logarithmic renormalization scale defined in terms of the renormalized mass, $m_r$, yielding
the inverse correlation length.
Eq. (\ref{Eq:RG-inv})
is an important result, since it allows one to study the critical topological behavior with a vanishing renormalized
Josephson coupling $J_r$, while still having $\theta\neq 0$, corresponding to the critical point of the topological phase 
transition. 

Combining both CS terms stemming from the two surfaces, we obtain that
the axion term in the bulk does not get renormalized either, so the result (\ref{Eq:RG-inv}) also holds in the bulk.
An interesting consequence of this analysis is that the RG flow of the bulk theory does not differ significantly from the one where
the axion term is absent, which is just given by the well-known analysis of Coleman and Weinberg.\cite{Coleman-Weinberg}
In this case, the phase transition is known to be of first-order, irrespective of the superconducting regime being of type I or type II.
However, this is not the case for the phase transition on the surfaces. Indeed, the
vanishing of Feynman graph of Fig. \ref{Fig:fish} when
$\theta\neq 0$ implies that a contribution $\sim \hat q^4$ is absent in the RG $\beta$
function of $\hat u$. Here are $\hat q=q_r/m_r$ and $\hat u=u_r/m_r$ renormalized
dimensionless couplings defined on the surface of a TSC. The one-loop RG functions
for a superconductor in 2+1 dimensions with a CS term, and $N$ complex order parameter
fields, are obtained in a way similar as in Ref. \onlinecite{Halperin-Lubensky-Ma},
except that we use the propagator (\ref{Eq:photon-prop})  with $m_A=0$ in the Feynman
diagrams involving photon lines. The result is,
\begin{equation}
\frac{d\hat q^2}{dl}=\left(\frac{N\hat q^2}{24\pi}-1\right)\hat q ^2,
\end{equation}
\begin{equation}
\label{Eq:betau}
\frac{d\hat u}{dl}=-\left[1+\frac{4}{3\pi}\frac{\hat q^2}{\left(1+\frac{\hat q^2|\theta|}{8\pi^2}\right)^2}\right]\hat u+\frac{(N+4)}{8\pi}\hat u^2,
\end{equation}
where we now have,
\begin{equation}
\theta=\sum_{j=1}^{N}C_{1j}\theta_j,
\end{equation}
where $C_{1j}$ are the Chern numbers associated to the helicity of the $N$ Fermi surfaces involved.\cite{Qi-Witten-Zhang}
Note that we have not expanded $1/(1+\hat q^2|\theta|/8\pi^2)$ in powers of $\hat q$, because $\theta$ can also be very large, so 
that the product $\hat q^2|\theta|$ is not necessarily small. 

It is easily seen that the above RG equations have an infrared stable fixed point for all $N$, in contrast to the analysis by
Halperin {\it et al.} for the non-topological superconductor, where infrared fixed points are only found for $N>183$.\cite{Note-HLM}
However, there is a stability condition involving $\theta$ that has to be fulfilled in the case of a TSC. It is obtained by considering the critical correlation function of the superconducting order field components,
\begin{equation}
\langle \phi_i(x)\phi_j^*(0)\rangle\sim\frac{\delta_{ij}}{|x|^{1+\eta}},
\end{equation}
where at one-loop,
\begin{equation}
\eta=-\frac{16}{N\left(1+\frac{3|\theta|}{\pi N}\right)^2},
\end{equation}
which implies the inequality $\eta>-1$. The latter inequality is fulfilled provided,
\begin{equation}
\label{Eq:ineq}
|\theta|>\frac{\pi}{3}(4\sqrt{N}-N),
\end{equation}
and we see that for $N\geq 16$ a quantum critical point is obtained for all values $\theta$, showing that at least sixteen Weyl fermions
are necessary to have a quantum critical point. Values of $\theta$ violating  the inequality $\eta>-1$ correspond to a situation where 
a continuum limit cannot be defined and is therefore unphysical.   
Thus, in order to have a physically meaningful phase transition, the inequality (\ref{Eq:ineq}) has to be satisfied. We note that
the lower bound for $|\theta|$ is larger than $\pi$ when $N=2$. Therefore, the TR symmetric
value $\theta=\pi$ obtained at the mean-field level does not produce a second-order phase transition when quantum fluctuations
are accounted for.  The one-loop RG predicts that a second-order phase transition occurs for $\theta=\pi$ only if $N\geq 10$.

It is tempting to relate the critical value of $N$ to the $Z_{16}$ classification.\cite{Wang-Senthil-Z16}
However, at this stage it would be too speculative, since
higher order results may affect the values of $N$ for which critical points obey the inequality $\eta>-1$.

As a final remark on the RG analysis, let us comment on another possible renormalization scheme allowing to continuously connect 
the cases $\theta=0$ and $\theta\neq 0$. The  infrared divergences stemming from the photon propagator for $\theta=0$ requires 
defining the renormalized four-point vertex at external nonzero momenta. This yields a renormalization scale $\mu$ that replaces $m_r$ 
in the RG flow. Since the diagram of Fig. \ref{Fig:fish} does not vanish for nonzero external momenta, it turns out that a $\hat q^4$ term 
would be generated in the RG $\beta$ function of $\hat u$, even when $\theta\neq 0$. This  $\hat q^4$ term would have 
a $\theta$-dependent coefficient allowing to smoothly connect the result to the known RG equations 
of a topologically trivial superconductor 
in the limit $\theta\to 0$. We would find once 
more that a second-order phase transition occurs for large enough $\theta$ and a stability criterion would follow  
from the inequality $\eta>-1$.\cite{deCalan} The latter inequality would yield in this case values of $\theta$ leading to 
 a {\it negative} coefficient of the $\hat q^4$ term, with values of $\theta$ yielding a positive coefficient of $\hat q^4$ violating the condition 
 $\eta>-1$. This implies that the RG $\beta$ function of $\hat u$ can actually not be continuously connected to 
 the $\theta=0$ regime, since to this end it would be necessary to enter a regime where the continuum limit is not even 
 defined. 

\subsection{Summary of the phase structure}

There are two important consequences of quantum fluctuations as unveiled by the analysis in this Section. First, we note that it is not
possible to reach the topologically trivial phase from the topologically nontrivial one within the RG. Simply taking the limit
$\theta\to 0$ does not recover the RG flow of topologically trivial superconductors, while this limit can be realized classically.  Second, and
most importantly, the phase transition in the bulk is always a first-order one, while a second-order phase transition is possible on
the surface. This is not the case for the topologically trivial superconductor, where a second-order phase transition on the surface is
only obtained for sufficiently large $N$.

\section{Vortex-free anomalous Hall effect}

We next turn to the Meissner effect aspects of a TSC, which as we will now show, implies an anomalous Hall effect even in the absence of vortices. This is more conveniently done by
rewriting the Lagrangian in a London limit exhibiting explicitly electric and magnetic
fields, i.e.,
\begin{eqnarray}
 \label{Eq:Seff}
 {\cal L}_{\rm eff}&=&\frac{1}{2}\left({\bf E}^2-{\bf B}^2\right)+\frac{e^2\theta}{4\pi^2}{\bf E}\cdot{\bf B}
 \nonumber\\
 &+&\frac{1}{2}\sum_{i=L,R}\rho_i^2(\partial_\mu\theta_i-q A_\mu)^2+J\rho_L\rho_R\cos\theta
 \nonumber\\
 &-&\frac{m^2}{2}(\rho_L^2+\rho_R^2)-\frac{u}{8}(\rho_L^2+\rho_R^2)^2.
\end{eqnarray}
From the effective Lagrangian we obtain that the electric displacement and magnetic fields are given respectively by
%
 ${\bf D}={\bf E}+e^2\theta{\bf B}/\pi$ and
%
 ${\bf H}={\bf B}-e^2\theta{\bf E}/\pi$,
while the superconducting  current is given by,
\begin{equation}
\label{Eq:current}
 {\bf j}=q[\rho_L^2(\nablab\theta_L-q{\bf A})+\rho_R^2(\nablab\theta_R-q{\bf A})].
\end{equation}
From Eq. (\ref{Eq:current}) we obtain the usual London equation in absence of vortices,
%
 $\nablab\times{\bf j}=-(1/\lambda^2){\bf B}$,
where $\lambda^2=1/m_A^2$ is the square of the penetration depth. Thus, the Maxwell equation in the presence of the axion field,
\begin{equation}
\label{Eq:Maxwell}
 \nablab\times{\bf B}={\bf j}+\partial_t{\bf E}
 +\frac{e^2}{\pi}(\nablab\theta\times{\bf E}+\partial_t\theta~{\bf B}),
\end{equation}
yields the  equation determining the London electrodynamics of the TSC in the form,
\begin{eqnarray}
 \label{Eq:London}
  \partial_t^2{\bf B}
  -\nabla^2{\bf B}+m_A^2{\bf B}
&=&\frac{e^2}{\pi}[\nablab\times(\nablab\theta\times{\bf E})
\nonumber\\
&+&\nablab\times(\partial_t\theta~{\bf B})].
\end{eqnarray}
%
For the axion field we obtain the equation of motion,
\begin{equation}
\label{Eq:SG}
\partial_t^2\theta-\nabla^2\theta+m_\theta^2\sin\theta
=\frac{e^2}{4\pi^2}\left(\frac{1}{\rho_L^2}+\frac{1}{\rho_R^2}\right){\bf E}\cdot{\bf B}.
\end{equation}

In the low frequency regime and in absence of vortices, the London equation (\ref{Eq:London}) simplifies to,
\begin{equation}
 \label{Eq:London-1}
 -\nabla^2{\bf B}+m_A^2{\bf B}=\frac{e^2}{\pi}[\nablab\times(\nablab\theta\times{\bf E})],
\end{equation}
while the current satisfies,
\begin{equation}
\label{Eq:London-j}
 -\nabla^2{\bf j}+m_A^2{\bf j}=-\frac{e^2m_A^2}{\pi}(\nablab\theta\times{\bf E}).
\end{equation}
The London equation for the electric field is unaffected by the axion term, retaining its traditional form, $-\nabla^2{\bf E}+m_A^2{\bf E}=0$. This result is closely related to the
fact that the electromagnetic energy density does not contain a magnetolectric term.

We now consider a solution with a simple geometry, namely, a semi-infinite TSC ($z\geq 0$)
with a surface at $z=0$ at an external electric field ${\bf E}_0=E_0\hat {\bf x}$ parallel
to the surface.  We obtain,
\begin{equation}
 \label{Eq:SG-1}
 -\frac{d^2\theta}{dz^2}+m_\theta^2\sin\theta=\frac{e^2}{4\pi^2}\left(\frac{1}{\rho_L^2}+\frac{1}{\rho_R^2}\right)E_x(z)B_x(z),
\end{equation}
where $E_x(z)=E_0e^{-m_Az}$, and
\begin{equation}
\label{Eq:London-Bx}
-\frac{d^2B_x}{dz^2}+m_A^2B_x=-\frac{e^2}{\pi}\frac{d}{dz}\left[E_x(z)\frac{d\theta}{dz}\right],
\end{equation}
\begin{equation}
\label{Eq:jx}
-\frac{d^2j_y}{dz^2}+m_A^2j_y=-\frac{e^2m_A^2}{\pi}E_x(z)\frac{d\theta}{dz}.
\end{equation}
The solution for Eq. (\ref{Eq:jx}) in terms of the axion is,
\begin{equation}
\label{Eq:sol-jy}
j_y(z)=\frac{e^2m_A^2 E_0}{\pi}\left[\frac{\theta(0)}{2m_A}e^{-m_Az}-e^{m_Az}\int_z^\infty dz'e^{-2m_Az'}\theta(z')\right],
\end{equation}
where $\theta(0)=\pi$.
Since ${\bf E}\cdot(\nablab\times{\bf j})=-\lambda^{-2}{\bf E}\cdot{\bf B}$, we obtain the following
relation,
\begin{equation}
\label{Eq:relation}
\frac{1}{\lambda^2}B_x(z)=m_Aj_y(z)+\frac{e^2m_A^2 E_0}{\pi}e^{-m_Az}[\theta(z)-\theta(0)],
\end{equation}
which implies $m_AB_x(0)= j_y(0)$. Thus, we find that the usual  boundary condition of the London theory,
%
 $dj_y/dz|_{z=0}=m_Aj_y(0)$,
is obviously fulfilled by the solution (\ref{Eq:sol-jy}) in the presence of the axion field. However, the
Maxwell equation (\ref{Eq:Maxwell}) in the static regime implies a boundary condition that deviates from the standard one in the London theory of non-topological superconductors,
\begin{equation}
 \label{Eq:BC-2}
 \left.\frac{dB_x}{dz}\right|_{z=0}=m_AB_x(0)+\frac{e^2E_0}{\pi}\left.\frac{d\theta}{dz}\right|_{z=0}.
\end{equation}

\begin{figure}
	\centering
	\includegraphics[width=9cm]{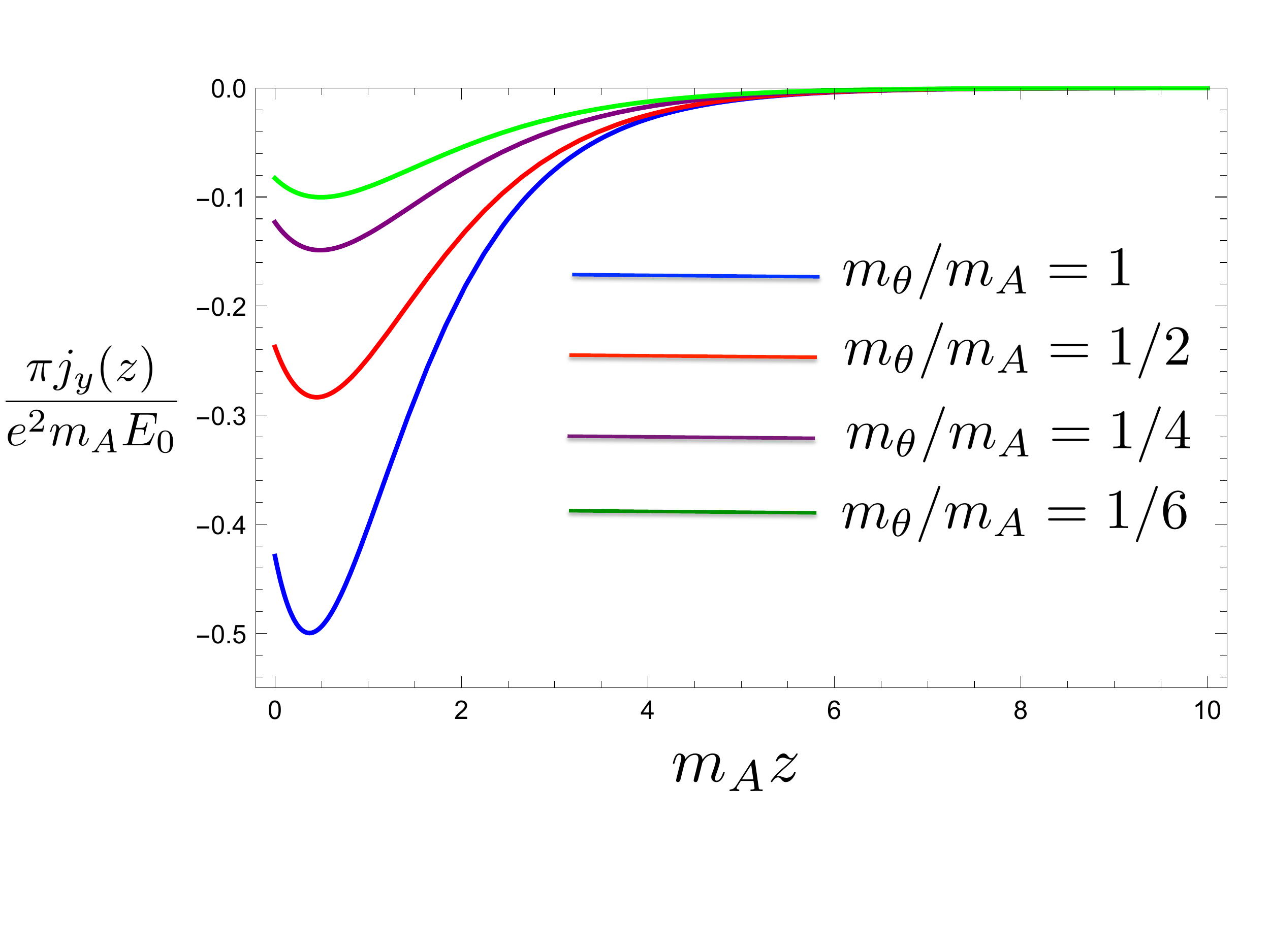}
	\caption{Induced anomalous Hall current $j_y(z)$ for $m_\theta/m_A=1,1/2,1/4,1/6$. }
	\label{Fig:anom-Hall}
\end{figure}

From Eqs. (\ref{Eq:sol-jy}) and (\ref{Eq:relation}) we see that an approximate solution can be obtained by considering
terms proportional to $e^4$ as being of higher order, which amounts to  approximating  Eq. (\ref{Eq:SG-1}) as being
homogeneous. In this case we can use the domain wall solution $\theta(z)=\pi+2\arcsin[\tanh(m_\theta z)]$ in Eq.
(\ref{Eq:sol-jy}), which yields $j_y(z)$ explicitly. The explicit solution for $j_y(z)$ with $m_\theta\neq m_A$
in terms of hypergeometric and Lerch transcedents, is not very illuminating.  Instead, we plot it
in Fig. \ref{Fig:anom-Hall} for four different values of the ratio $m_\theta/m_A$. It has a
negative sign, just like in the case of the anomalous Hall effect in high-$T_c$ superconductors
\cite{Anom-Hall-Effect-high-Tc}. As emphasized in the introductory paragraphs, the anomalous Hall
effect in non-topological superconductors has a quite different origin from the one discussed here.  {\it In
three-dimensional TSCs the anomalous Hall current arises independently of vortex motion and is associated with a
dissipationless current. }

For $m_\theta=m_A$ (blue curve in Fig. \ref{Fig:anom-Hall}), the expression for $j_y(z)$ does not involve special functions, reading,
\begin{eqnarray}
j_y(z)&=&\frac{e^2m_AE_0}{\pi}\left\{(\pi/2)e^{-m_Az}-2
\right.\nonumber\\
&-&\left. 2[e^{-m_Az}\arctan(e^{m_Az})-e^{m_Az}\arctan(e^{-m_Az})]\right\}.
\end{eqnarray}

In order to make connection with the quantization of Hall conductivity in the normal state,
it is instructive to integrate the current density over $z\in[0,\infty)$ to obtain the surface current density,
\begin{eqnarray}
j_y^{\rm surf}&= &\int_0^\infty dzj_y(z)=\frac{e^2}{2\pi}E_0\theta(0)
\nonumber\\
&-&\frac{e^2E_0m_A}{\pi}\int_0^\infty dz(e^{-m_Az}-e^{-2m_Az})\theta(z).
\end{eqnarray}
We obtain,
\begin{eqnarray}
j_y^{\rm surf}&= &\frac{e^2E_0}{2\pi}\left[2\psi\left(\frac{m_\theta+m_A}{4m_\theta}\right)-\psi\left(\frac{m_\theta+2m_A}{4m_\theta}\right)
\right.\nonumber\\
&+&\left. \psi\left(\frac{3m_\theta+2m_A}{4m_\theta}\right)-2\psi\left(\frac{3m_\theta+m_A}{4m_\theta}\right)
\right],
\end{eqnarray}
where $\psi(z)=\Gamma'(z)/\Gamma(z)$ is the digamma function. The normal state corresponds to $m_A/m_\theta\to 0$, which yields
$j_y^{\rm surf}=-(e^2/2)E_0$, i. e., the half-quantum of the quantized Hall conductivity.

\section{Conclusion}

In conclusion, we have shown that due to quantum electromagnetic fluctuations, the Higgs mechanism in three-dimensional TSCs implies a robust topological state of matter, since its RG flow
cannot be continuously deformed into the RG flow of a topologically trivial one.
This is an example of a topological state that is protected due to the coupling of phase and electromagnetic fluctuations via the axion term, with TR symmetry not being required.
In fact, TR  can be spontaneously broken by quantum fluctuations.
In this context, we have also shown that  a second-order quantum phase transition happens on the surface of a TSC, while its bulk undergoes a first-order phase transition. Without
the axion term a first-order phase transition would happen both in the bulk and on the surface, provided the superconductor is in the type I regime.

Another aspect of the Higgs mechanism we have studied is the influence of the axion term in
the Meissner effect. We have found that the gradient of the axion field on the surface
induces a transverse supercurrent. In the low frequency limit this implies a London regime
leading to the generation of an anomalous Hall current with a negative sign. This anomalous
Hall current is dissipationless and is the consequence of a Lorentz-like force involving
the relative superfluid velocity which is simply given by the gradient of the phase difference between the chiral superconducting components.

\acknowledgments

F.S.N. and I.E. acknowledge the Deutsche Forschungsgemeinschaft (DFG) for the financial support via the collaborative research
center SFB TR 12. A.S. acknowledges support from the Research Council of Norway, Grant Nos. 205591/V20 and 216700/F20, as well
as support from COST Action MP-1201 "Novel Functionalities Through Optimized Confinement of Condensates and Fields".
IE also acknowledges the financial support of the Ministry of
Education and Science of the Russian Federation in the framework of Increase Competitiveness Program of
NUST MISiS (N 2–2014–015).


\begin{thebibliography}{100}
\bibitem{Note-HLM} Note that in Ref. \onlinecite{Halperin-Lubensky-Ma} $N$ denotes the number of real components of the
order parameter, while we use $N$ as the number of complex components.
\end{thebibliography}
\end{document}